# Interpretable Data-Driven Demand Modelling for On-Demand Transit Services

**Nael Alsaleh***
Laboratory of Innovations in Transportation (LiTrans)
Department of Civil Engineering,
Ryerson University, Toronto, ON Canada, M5B 2K
Email: nael.alsaleh@ryerson.ca

**Bilal Farooq**
Laboratory of Innovations in Transportation (LiTrans)
Department of Civil Engineering,
Ryerson University, Toronto, ON Canada, M5B 2K
Email: bilal.farooq@ryerson.ca

**Abstract**

In recent years, with the advancements in information and communication technology, different emerging on-demand shared mobility services have been introduced as innovative solutions in the low-density areas, including on-demand transit (ODT), mobility on-demand (MOD) transit, and crowdsourced mobility services. However, due to their infancy, there is a strong need to understand and model the demand for these services. In this study, we developed trip production and distribution models for ODT services at Dissemination areas (DA) level using four machine learning algorithms: Random Forest (RF), Bagging, Artificial Neural Network (ANN) and Deep Neural Network (DNN). The data used in the modelling process were acquired from Belleville's ODT operational data and 2016 census data. Bayesian optimization approach was used to find the optimal architecture of the adopted algorithms. Moreover, post-hoc model was employed to interpret the predictions and examine the importance of the explanatory variables. The results showed that the land-use type was the most important variable in the trip production model. On the other hand, the demographic characteristics of the trip destination were the most important variables in the trip distribution model. Moreover, the results revealed that higher trip distribution levels are expected between dissemination areas with commercial/industrial land-use type and dissemination areas with high-density residential land-use. Our findings suggest that the performance of ODT services can be further enhanced by (a) locating idle vehicles in the neighbourhoods with commercial/industrial land-use and (b) using the spatio-temporal demand models obtained in this work to continuously update the operating fleet size.

**Keywords:** On-demand service (ODT); Demand modelling; Machine learning; Model interpretability; Shared mobility; Trip generation and distribution.

*\* Corresponding author*

# 1. Introduction

According to census 2016, the Greater Toronto and Hamilton Area (GTHA) has a population of more than 6.9 million and is considered as the most populous region in Canada. Approximately 43% of its population lives in dense-urban areas, while the rest in low-density areas, which represent 96% of the total land area of the GTHA. Similar pattern can be observed in the other part of the country. In such low-density and low-demand areas, fixed-route public transit service (FRT) operates with a low frequency and poor spatial coverage to compromise between the occupancy of the vehicles and their operational cost (Papanikolaou et al., 2017; Navidi et al., 2018; Franco et al., 2020). These conditions can render the FRT inconvenient and unattractive for residents. Thus, there is a strong need for more innovative and sustainable alternatives for the existing FRT services in order to mitigate traffic congestion, reduce emissions, as well as increase the efficiency and the attractiveness of public transit services in the low-density and low-demand areas. In recent years, with the advancements in information and communication technology, different emerging on-demand shared mobility services have been introduced as innovative solutions in the low-density areas, including on-demand transit (ODT), mobility on-demand (MOD) transit, and crowdsourced mobility services (Ma et al., 2020).

ODT is a dedicated fleet mobility service that is primarily used as an alternative for the FRT service in the low-demand areas using the existing infrastructure (buses and stops) and fare system. The ODT concept is to follow users' schedules and routes to reduce the operational cost of the service. (Papanikolaou et al., 2017; Sanaullah et al., 2021). Several municipalities in Canada have already replaced their FRT with ODT services, like Regina and Saskatoon in Saskatchewan, St. Albert, Calgary, Cochrane, and Okotoks in Alberta, as well as Chatham-Kent, Sault Sainte Marie, and Belleville in Ontario (Klumpenhouwer et al., 2020).

MOD transit is an integration between the FRT service and subsidized on-demand mobility services, like ridesourcing and microtransit services (Yan et al., 2019; Zhao et al., 2019). The FRT service operates along the major roads, where the trip demand is relatively high and travel distances are longer. On the other hand, the subsidized crowdsourced or dedicated fleet based on-demand mobility services are used in the low-demand areas to provide a first/last mile service or to act as a feeder to the FRT service. MOD service has been deployed in several cities in North America, including Los Angeles in California, King County in Washington, as well as Columbus, Gahanna and New Albany in Ohio. However, the use of crowdsourced mobility services is not limited to providing a first/last mile service in the low-demand areas, but it is also used as the main transit service in the Town of Innisfil, in Ontario. In 2017, the municipality of Innisfil created a partnership with Uber to provide subsidized ridesourcing service for residents instead of running FRT service. The main idea behind this partnership was to offer residents with an affordable, convenient, and cost-efficient transportation option (Zenasni, 2019).

Figure 1 illustrates the differences between transit services in low-density and low-demand areas, in terms of fleet ownership, vehicle size, supply resources, and level of sharing. In both ODT and FRT service, vehicles have a high capacity and owned by an organization (e.g., municipality), fleet size is dedicated, and the level of sharing is high. While in ridesourcing services, vehicles have a low capacity and owned by people, fleet size is non-dedicated, and the level of sharing is low. On the other hand, MOD transit system is a combination between FRT and ridesourcing services.

A well planned and designed implementation of the on-demand shared mobility services in low-density and low-demand areas can have a wide range of positive effects. These services offer transit operators a new opportunity to deliver sustainable, convenient, attractive, and cost-efficient transit services. Moreover, they allow public transit users better access to jobs, engagement in various activities, and maintaining of social relationships. However, the performance of these services is highly affected by the number of requests the system receives and their distribution over space and time (Navidi et al., 2018). Hence, it is of vital importance to understand and model the demand for the emerging shared mobility services in low density areas. The demand modelling related studies are limited to crowdsourced mobility services. As far as we know, there are no studies that have developed demand models for on-demand dedicated fleet mobility services (i.e., ODT).

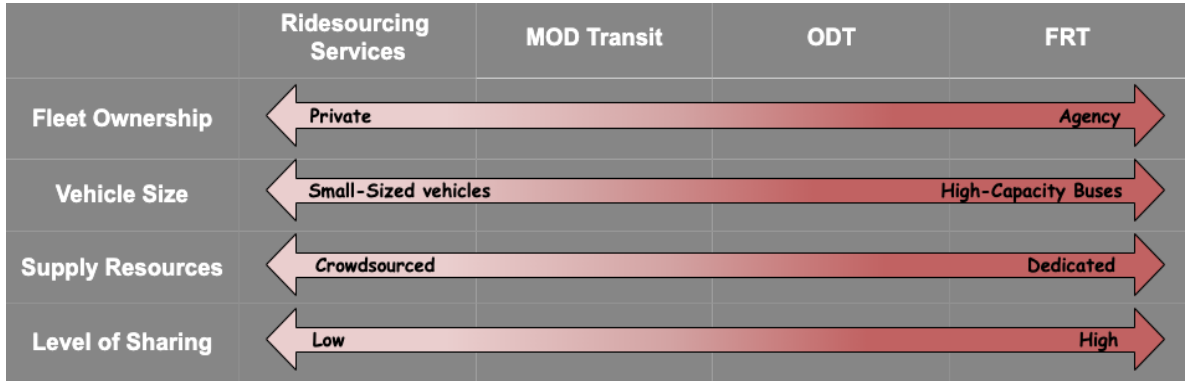

**Figure 1.** Transit services in low-density and low-demand areas

This study aims to fill the gap in the literature by developing trip production and distribution models, the first two steps of the classic four-step model, for ODT services at dissemination area level. The aim of these models is twofold: (a) predicting ODT daily trip production and distribution levels as well as (b) investigating the effect of demographic characteristics and trip characteristics on the demand levels. Two main approaches have been adopted in the literature for the development of trip production and distribution models: machine learning algorithms, like artificial neural network and random forest, and traditional statistical methods, like gravity models. However, in comparison with traditional statistical methods, machine learning algorithms are more flexible, more accurate, and require fewer parameters to be estimated (Pourebrahim et al., 2019; Tillema et al., 2006). Thus, in this study, both the trip production and distribution models were developed using four machine learning algorithms, namely Random Forest (RF), Bagging, Artificial Neural Network (ANN), and Deep Neural Network (DNN). The data used in the modelling process are acquired from Belleville's ODT operational data and 2016 census data. Moreover, SHAP (SHapley Additive exPlanation), a post-hoc model analysis method, was employed to interpret the predictions and examine the importance of the explanatory variables. Based on our findings, we further provide some useful policy recommendations to operators and municipalities for sustainable planning, design, and operation of new and ongoing ODT projects.

The rest of the paper is organized as follows. Section 2 presents the existing demand modelling studies for on-demand shared mobility services. Section 3 provides the main demographic and public transit information in the city of Belleville. The framework used in this study and the specifications of the adopted machine learning algorithms are explained in Section 4. The results of the trip production and distribution models are presented in Section 5. Section 6 provides a detailed discussion and some policy recommendations drawn from our results. Finally, the paper ends with a conclusion and future work.

## 2. Background

This section reviews the previous studies that modelled and analyzed the demand for on-demand shared mobility services. It is important to note that there is an obvious lack of research conducted on predicting the demand for ODT services, which may be because of the lack of readily available data on the ODT services.

Wang et al. (2014) used a multivariate analysis to investigate the impact of the socioeconomic characteristics on the demand of the microtransit service for the metropolitan region of Greater Manchester. The study found that the areas that have low population density, auto-ownership percentages, income, employment, and education tend to have higher demand. Jain et al. (2017) reviewed the existing literature on the usability of microtransit services. The study identified 11 factors to affect the usage of the service linked to the socioeconomic characteristics of users, demographic characteristics of the service area, as well as the level of service attributes. These factors include age, gender, purpose of the trip, availability of the driving licence, the auto-ownership percentages, income, average household size, availability of train stations, and average trip waiting and walking times. Sanaullah et al. (2021) performed a detailed spatio-temporal demand and supply analysis for Belleville's ODT service based on the data collected from September 2018 till May 2019. The study also investigated the impact of population density, median income, and working-age percentages levels on the ODT trip production and attraction levels using GIS and k-means clustering technique. The results revealed that the dissemination areas with higher population density, lower median income, or higher working-age percentages are associated with higher ODT trip attraction levels.

Young and Farber (2019) used 2016 Transportation Tomorrow Survey (TTS) to describe the characteristics of ridesourcing users and capture the effect of ridesourcing services on the ridership of other travel modes in Toronto, Canada. The study found that most of the users were young, wealthy, and full-time workers. Moreover, the results revealed that ridesourcing services were mainly used to return home as well as to engage in non-work-related activities. Alemi et al. (2018) developed a binary logit model to examine the factors affecting the adoption of ridesourcing services among young adults and middle-aged groups in California. The results indicated that older millennials, highly educated travellers, app-based transportation services' users, and those who used taxi and carsharing services before are more likely to adopt the service. Another study by the same authors revealed that the sociodemographic factors have a significant impact on the adoption of the ridesourcing services. Moreover, the study showed that the usage frequency of the ridesourcing services is affected positively by the activity density and negatively by the land-use mix (Alemi et al., 2019). In the study conducted by Marquet (2020), the actual ridesourcing data were used to investigate the relationship between the spatial distribution of the trips and the neighborhood characteristics in the city of Chicago. The results revealed that areas with higher population density, income, or employment percentages were associated with higher ridesourcing trips. Besides, areas with poor transit connectivity were found to generate and attract more trips. Another recent study in Chicago, Illinois, revealed that areas with higher incomes, higher percentage of workers, less private vehicle ownership, higher population and employment densities, higher land-use diversity, and fewer parking spots, were found to have higher ridesourcing trips (Ghaffar et al., 2020).

Trip-based travel demand modelling approach is simple, easy to use, requires manageable data, and less expensive for small towns to develop and use compared with the activity-based travel demand modelling approach. Hence, several researchers have adopted this approach to develop travel demand models for ridesourcing services using machine learning algorithms. Ye (2019) developed a trip production model for ridesourcing services using operational data collected from Uber service in Washington D.C. and New York city from April 2015 to September 2017. For this reason, the author used four machine learning methods, namely regularized regressions, trees, artificial neural network, and auto regressive integrated moving average (ARIMA). The results showed that the LASSO algorithm (one of the regularized regressions' algorithms) outperformed the other algorithms in predicting the pickup demand. In the study conducted by Saadi et al. (2017), a short-term spatio-temporal demand model was developed to predict the demand of ridesourcing services in China. The authors used a single decision tree, Bagging, random forest (RF), boosted decision trees, and artificial neural network algorithms to predict the demand. The results revealed the boosted decision trees algorithm had the highest predictive accuracy among the other algorithms in predicting the demand. In another study, a wave support vector machine (Wave SVM), support vector machine (SVM) and ARIMA algorithms were adopted to predict the demand of DiDi ridesourcing service in China. The results showed that the Wave SVM method outperforms the other two algorithms in predicting the demand (Li et al., 2017). In the study by Yan et al. (2020), random forest (RF) and traditional multiplicative models were adopted to develop a trip demand model for ridesourcing services at census tract (CT) level in Chicago. The results showed that RF algorithm had a higher predictive accuracy compared to the traditional multiplicative models. Moreover, the demand values were most affected by the socioeconomic and demographic variables.

Several researchers have used deep learning approach, a branch of machine learning, to obtain more complicated and accurate models for ridesourcing services. For instance, Chao et al. (2018) developed a trip production model for Uber service using a Long Short-Term Memory (LSTM) Networks model, time-varying Poisson model and regression tree model. The results showed that the LSTM model outperformed the other two models in predicting the demand. Lan (2020) applied historical models, multi-layer perceptron (MLP), convolutional neural network (CNN) and convolutional Long Short-Term Memory (ConvLSTM) algorithms to predict the demand of DiDi ridesourcing service in China. The results revealed that ConvLSTM had the highest predictive accuracy among the other models in predicting the demand. Wang et al. (2019) developed a short-term trip production model for DiDi ridesourcing service in Chengdu City, China, using CNN and LSTM algorithms. The results showed that the CNN model outperformed the LSTM model in predicting the demand.

It can be noted that the existing literature have not (a) identified the relative importance of the area wide characteristics, (b) examined the impact of trip characteristics, or (c) developed demand models for on-demand dedicated fleet mobility services. Therefore, this study aims to bridge these research gaps by developing trip production and distribution models for ODT services based on demographic characteristics and trip characteristics. Moreover, the impact of these characteristics on the ODT demand levels and their relative importance are also carried out in this study.

## 3. Case Study

The City of Belleville, Ontario, introduced one of the first ODT services in September 2018, where a late-night fixed-route public transit line was converted to an on-demand service as a pilot project. In this study, we use the operational data of Belleville's ODT service as a case study to develop trip production and distribution models for ODT services. This section provides a brief overview of the main demographic information and public transit service in the city of Belleville.

Belleville is a city in Eastern Ontario, Canada, located approximately 190 km east of Toronto and it is the largest urban centre in Quinte region (see Figure 2). The city has an area of 246.8 square km, with a population density of 205.1 people per square km. According to the 2016 census data, Belleville has a population of 50,716 people, around 63% of them are in the working age group (15 to 64 years old). The median household income (after-tax) in Belleville is 53,367 CAD, which is a fair bit lower than the national average at 61,348 CAD. Belleville's public transit service consists of 10 routes covering the urban area of the city with a total fleet of 16 buses. In January 2018, the city introduced a late-night fixed-route public transit line (Route 11) to satisfy shift workers travel needs at night. The fixed service was operated with two 40-ft buses, 30 minutes headway, and from 9:30 PM to 12:30 AM. However, to minimize the operational cost of the service and increase its efficiency, it was converted to an on-demand service as a pilot project in September 2018 (Pantonium, 2019; Sanaullah et al., 2021).

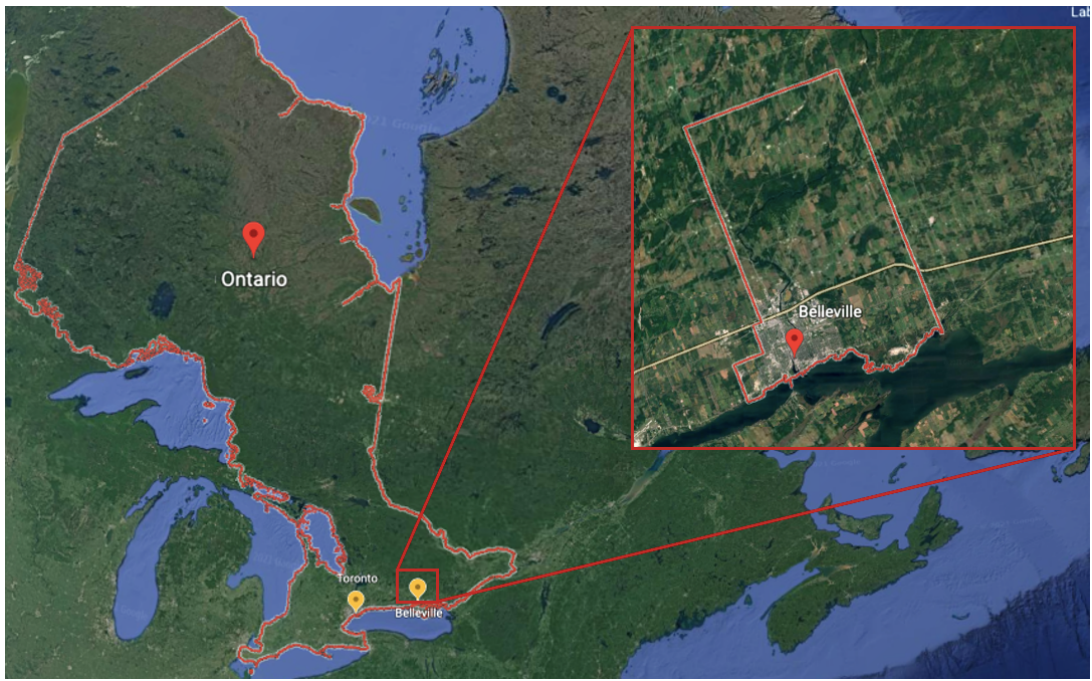

**Figure 2.** City of Belleville

The ODT service is using the same buses (40-ft), stops, and fares (3 CAD) as used for the fixed-route service. Unlike the late-night fixed-route service, ODT covers the entire city stops based on users' requests (see Figure 3). Besides, it continuously updates its routes, timetable, pickup and drop off locations according to the real-time spatio-temporal demand the system receives. The ODT service operates from 9:00 PM to 12:00 AM during the week and from 7:00 PM to 12:00 AM on the weekends. Users can request a trip through a mobile app called "*On Demand Transit Rider App*" or through phone. While making a trip request, the user is asked to provide information about the pickup and dropoff locations, the time for pickup, the tolerance time, and the number of riders to be picked up. However, users who are not familiar with smartphones or do not have access to internet can use the service by just showing up at the stop when the ODT bus arrives.

Converting the late-night fixed route service in Belleville to an on-demand service have caused a positive change in travel patterns and a sustained increase in demand levels (Sanaullah et al., 2021). According to the authors, this could be due to the features, like the flexibility in the pickup and dropoff locations and the larger coverage area. Therefore, in this study, we develop trip production and distribution models for Belleville's ODT service based on the operational data collected from September 18, 2018 to May 29, 2019. The dataset contains detailed information about users (ID, requested pickup time, actual pickup time, origin and destination stops), trips (date, origin and destination stops, status), journeys (ID, longitude and latitude, time stamp), and buses (operating hours per day, start and end locations). For further details about the ODT service in Belleville and its operational data we refer readers to Sanaullah et al. (2021) and the official city website at: https://www.belleville.ca/en/walk-ride-and-drive/transit.aspx.

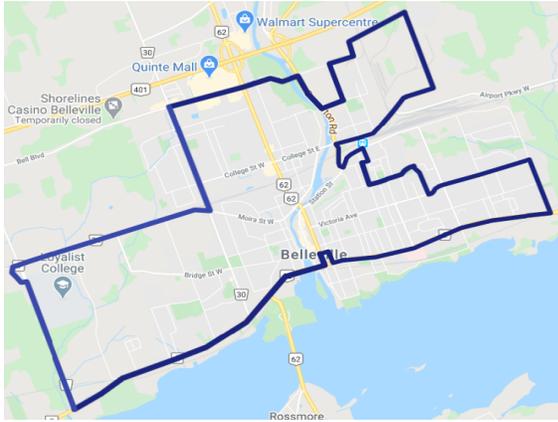 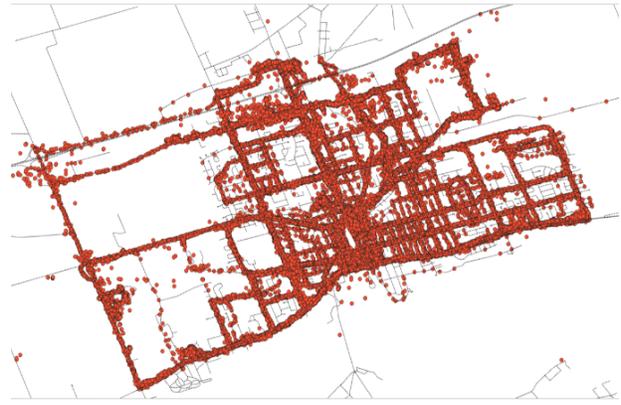

a) Fixed service                                b) On-demand service

**Figure 3.** Belleville public transit fixed-route and on-demand services

## 4. Methodology and Model Specifications

Four machine learning algorithms were used in the current study to develop trip production and distribution models for Belleville's ODT service, namely random forest (RF), Bagging, artificial neural network (ANN), and deep neural network (DNN). Figure 4 illustrates the framework used to develop the models. The modelling process consists of two main parts: selection of the variables and the specifications of the models. In this section, we discuss these parts in detail.

4.1. Data Preparation

In this study, the data were acquired from (a) Belleville's ODT operational data collected from September 18, 2018 till May 29, 2019 and (b) 2016 census data. The operational data were used to obtain the following information: (a) pickup and dropoff locations for each ODT trip record, (b) trip date (day and month), and (c) the daily hours of operation. Trip origins and trip destinations were aggregated in each DA on each day to form the dependent variables for the trip production and distribution models. Then, the daily number of trips produced from each DA was used as a dependent variable for the trip production model. While the daily number of trips between DAs was used as a dependent variable for the trip distribution model. On the other hand, trip date and daily hours of operation information were used as explanatory variables in both models to explain the temporal variations in the ODT demand. Thus, the trip date information was represented by two discrete variables: day-of-week and month-of-year variables. These variables were defined to explain the daily and monthly variations in demand. Furthermore, as pointed out in Section 3, the ODT service in Belleville operates for 3 hours during the weekdays and 5 hours during weekends. Accordingly, the daily hours of operation information was defined as a binary variable to capture the difference between weekday and weekend demand.

Moreover, demographic characteristics are important factors in explaining the demand of on-demand shared dedicated fleet mobility services. Previous literature identified several demographic characteristics to affect the demand of on-demand mobility services, including population density, gender, median income, average household size, working age percentages, auto-ownership percentages, education, and the availability of driving licence (Wang et al., 2014; Jain et al., 2017; Sanaullah et al., 2021). However, the auto-ownership percentages and the availability of driving licence information are not available in 2016 census data, while the distribution of the education categories' values is homogenous among Belleville's dissemination areas. Therefore, in the current study, the population density, median income, working age percentages, average household size, and gender (percentage of males) characteristics were used as explanatory variables to explain the spatial variations in the ODT demand. The demographic variables were used for the trip origin in the trip production model and for both trip origin and destination in the trip distribution model. Hence, 8 explanatory variables were used for the ODT trip production model and 13 explanatory variables for the ODT trip distribution model. Table 1 provides descriptive statistics for both the explanatory and dependent variables.

According to Table 1, the ODT daily trip production variable has a minimum value of 1 trip/DA, a maximum value of 35 trips/DA, and a mean value of 4.3 trips/ DA. On the other hand, the ODT trip distribution variable has a minimum value of 1 trip/DA, a maximum value of 11 trips/ DA, and a mean of 2.4 trips/DA. Based on these values, it can be noticed that the ODT daily trip demand is very low, when compared with the daily demand of ridesourcing services seen in the literature. Additionally, the frequency of the low demand values is much higher than the frequency of the higher demand values. This is expected since, as explained in Section 3, the service is used in a low-density and low-demand area and operates only for a limited number of hours per day. However, due to the low demand values, we modelled the daily demand with classification algorithms. In light of this, the daily trip production and distribution values were further categorized into k predefined clusters using k-means clustering algorithm and elbow method, such that each cluster represents a certain demand level.

K-means algorithm is a simple, but powerful and the most widely used clustering technique. The algorithm is used to categorize unlabelled dataset into predetermined number of clusters in such a way that the data points that belong to the same group have similar characteristics. However, the algorithm requires user to specify the number of clusters beforehand (Celebi et al., 2013; Capó et al., 2020). Elbow method is a heuristic used to obtain the optimum number of clusters based on the total within-cluster sum of squares value (distortion score). The optimum number of clusters (K) is defined as the number of clusters after which the slope starts to level out (Nainggolan et al., 2019; Bholowalia and Kumar, 2014). K-means algorithm and elbow method have been widely used in transportation related studies. For instance, they were used to identify activities (Poucin et al., 2018), obtain the virtual station of bike aggregation (Hua et al., 2020), and determine trip attraction levels for ODT service (Sanaullah et al., 2021).

Figure 5 presents the clustering analysis results for the daily trip production values. From Figure 5a, it is observed that the daily trip production values can be best represented by 3 clusters, as the distortion score does not decrease considerably after this value. The first cluster contains daily trip production values of 1 trip/DA, the second cluster contains the values of 2 to 5 trips/DA, and the values of more than 5 trips/DA are included in the third cluster (Figure 5b). We refer to the first, second, and third clusters respectively as the low, medium and high trip production levels. As Figure 5c depicts, the low trip production level has the highest frequency and accounts for 40% of the total daily trip production values, followed in order by the medium (36%) and the high (24%) daily trip production levels.

Similarly, the clustering analysis results for the daily trip distribution values are illustrated in Figure 6. It can be seen that the distortion score decreases significantly until K=3 and goes down slowly after this point, indicating the location of the optimal number of clusters (Figure 6a). Thus, the daily trip distribution values were grouped into 3 categories as follows (Figure 6b): low trip distribution level (1 trip/day), medium trip distribution level (2 trips/day), and high trip distribution level (more than 2 trips/day). As Figure 6c presents, the high trip distribution level has the highest frequency and accounts for 39% of the total daily trip distribution values, followed by medium (37%) and low (24%) levels. It is worth mentioning that increasing or decreasing the number of clusters used for the daily trip production and distribution variables would affect the modelling results negatively. Increasing the number of clusters will render the dataset highly imbalanced which, in turn, will reduce the predictive accuracy of the models. On the other hand, decreasing the number of clusters will reduce the explanatory power of the models.

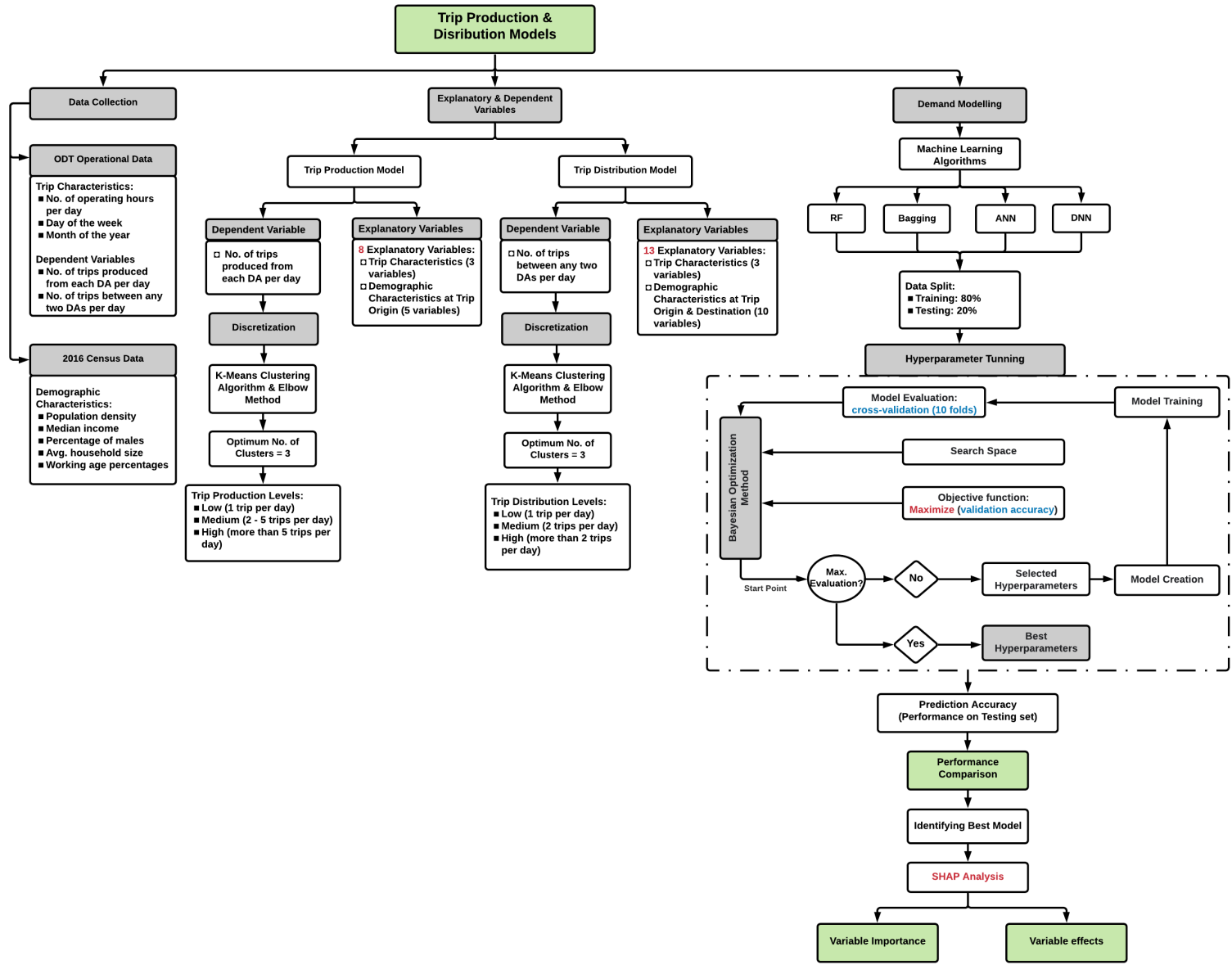

**Figure 4.** Demand modelling framework for ODT services

Table 1. Descriptive statistics for the explanatory and dependent variables

| Variable | Source | Unit | Type | Min | Max | Mean | Std. Dev. | Description |
|---|---|---|---|---|---|---|---|---|
| **Dependent Variables** | | | | | | | | |
| Number of trips produced from each DA per day | BODT[1] | Count | Discrete | 1 | 35 | 4.3 | 5.29 | - |
| Number of trips between any two DAs per day | BODT | Count | Discrete | 1 | 11 | 2.4 | 1.18 | - |
| **Explanatory Variables** | | | | | | | | |
| **Demographic Characteristics for Origin** | | | | | | | | |
| Population Density | 2016 Census | People/km$^2$ | Continuous | 63.2 | 8,139.8 | 1,700.7 | 1,628.02 | - |
| Median income | 2016 Census | CAD | Continuous | 24,640 | 87,296 | 49,506.2 | 13,522.32 | - |
| Average household size | 2016 Census | Count | Continuous | 1.5 | 2.8 | 2.2 | 0.30 | - |
| Percentage of males | 2016 Census | % | Continuous | 37.2 | 56.8 | 47.2 | 4.24 | - |
| Working age percentages | 2016 Census | % | Continuous | 38.9 | 80.7 | 63.5 | 8.52 | - |
| **Demographic Characteristics for Destination** | | | | | | | | |
| Population Density | 2016 Census | People/km$^2$ | Continuous | 63.2 | 8,139.8 | **2,075** | 2,094.04 | - |
| Median income | 2016 Census | CAD | Continuous | 24,640 | 87,296 | 46,851.6 | 13,407.37 | - |
| Average household size | 2016 Census | Count | Continuous | 1.5 | 2.8 | 2.17 | 0.25 | - |
| Percentage of males | 2016 Census | % | Continuous | 37.2 | 56.8 | 47.5 | 3.93 | - |
| Working age percentages | 2016 Census | % | Continuous | 38.9 | 80.7 | 66.4 | 6.54 | - |
| **Trip Characteristics** | | | | | | | | |
| Hours of operation per day | BODT | Count | Binary | 0 | 1 | - | - | 0: if 3 hours 1: if 5 hours |
| Day of the week | BODT | - | Discrete | 1 | 7 | - | - | Days from Saturday to Friday |
| Month of the year | BODT | - | Discrete | 1 | 9 | - | - | Months from September to May. |

1: BOD = Belleville's ODT Operational Data

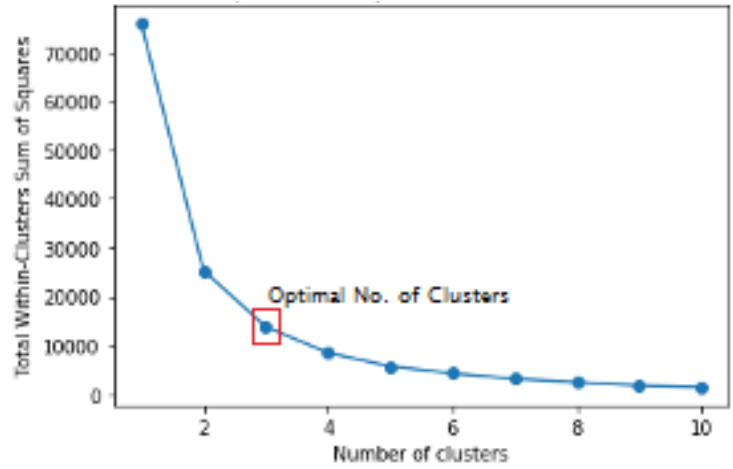

a) Optimal number of clusters

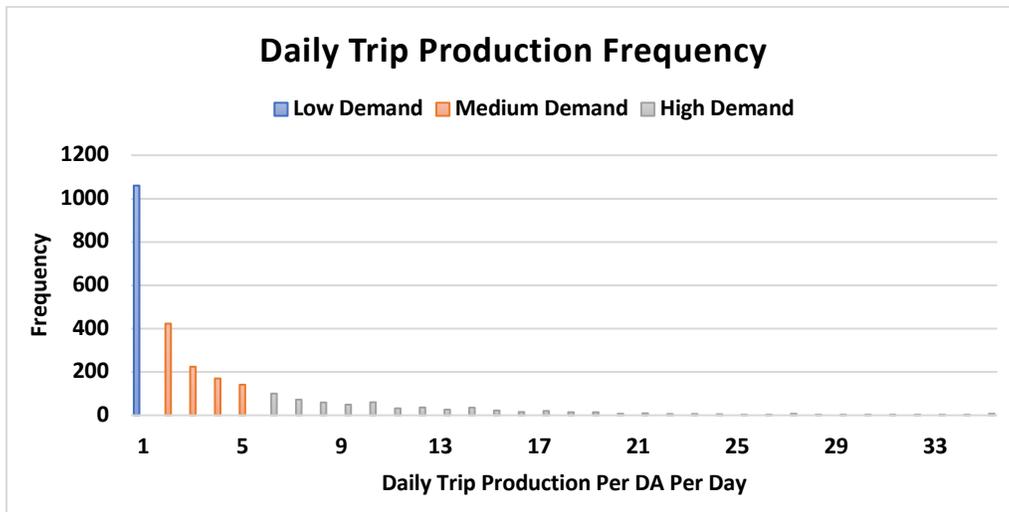

b) Daily trip production frequency

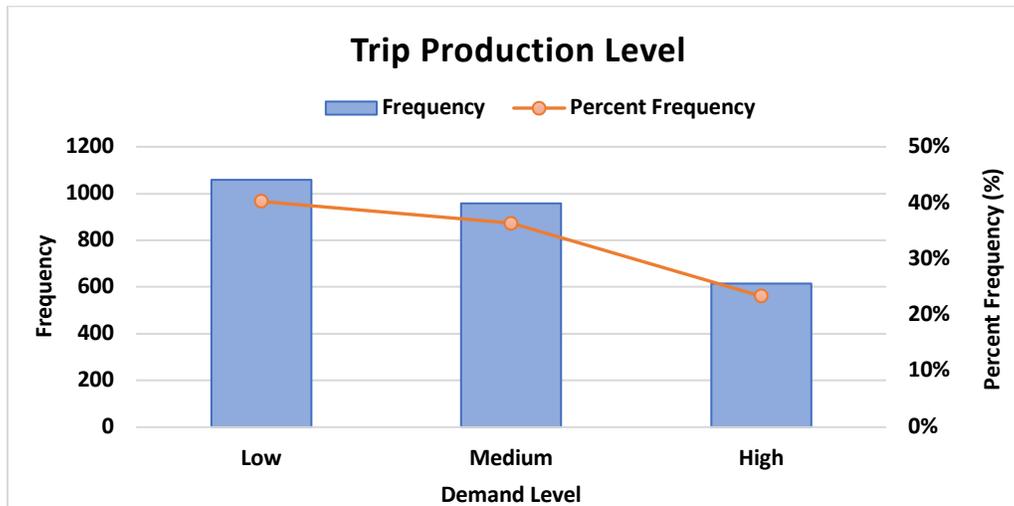

c) Demand levels frequency

**Figure 5.** Trip production (a) optimum number of clusters, (b) frequency, and (c) levels

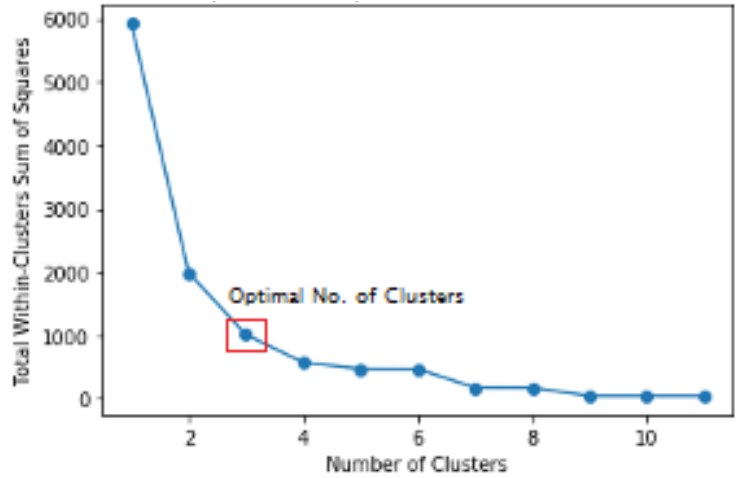

a) Optimal number of clusters

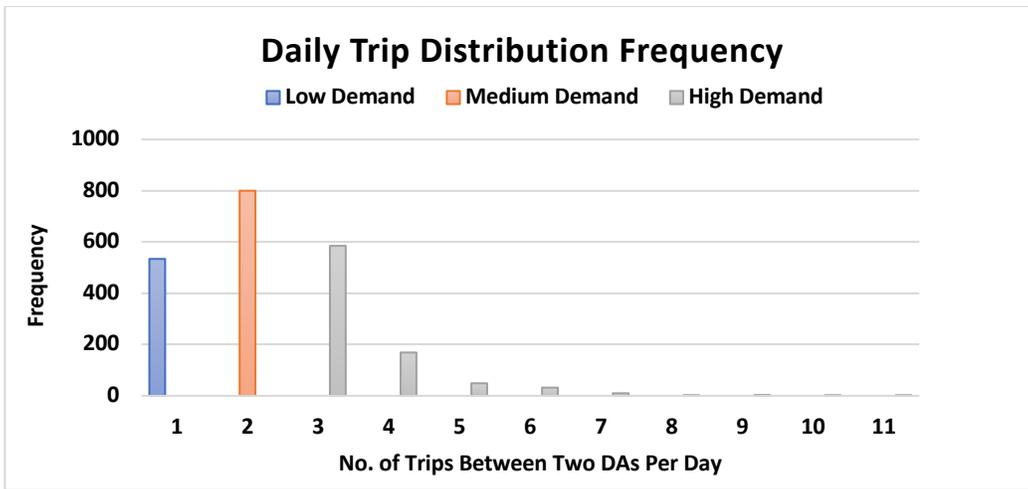

b) Daily trip distribution frequency

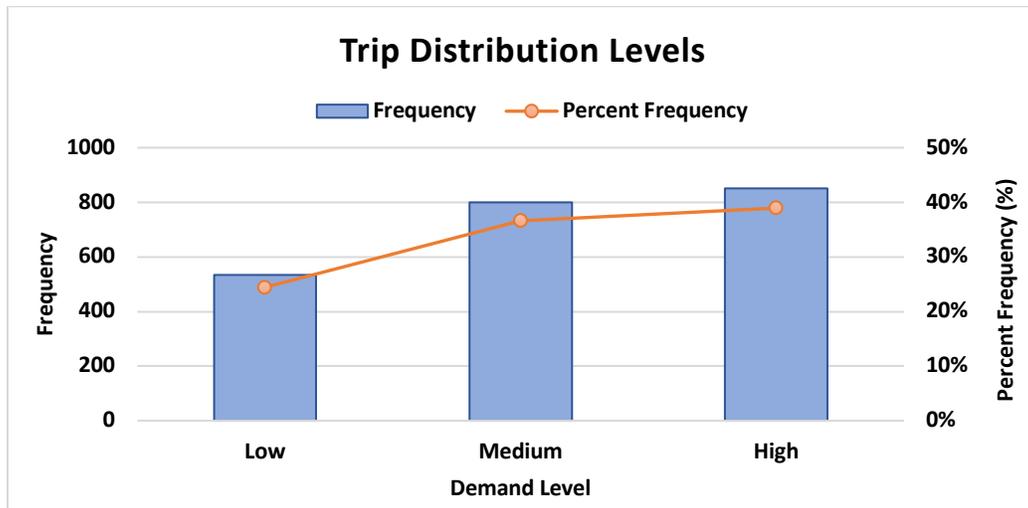

c) Demand levels frequency

**Figure 6.** Trip distribution (a) optimum number of clusters, (b) frequency, and (c) levels

## 4.2 Model Specification

In the current study, RF, Bagging, ANN, and DNN algorithms were used to predict the ODT daily trip production and distribution levels. Neural networks are a set of interconnected artificial neurons designed to perform several tasks such as clustering, classification, predictions, and optimization (Chen et al., 2008). They have been extensively used in various applications and areas, due to their ability to capture complex and nonlinear relationships. On the other hand, RF and Bagging (also known as a bootstrap aggregating) are powerful ensemble algorithms used primarily for classification and regression analysis. Both algorithms build multiple classifiers and aggregate them together to obtain more consistent and accurate predictions. However, they differ in the way the base classifiers are build. In RF algorithm, a random subset of features is considered when splitting nodes, while in Bagging algorithm, all features are used to split each node (Gislason et al., 2006).

The performance of machine learning algorithms depends on their hyperparameter settings. However, finding the optimal architecture for a particular machine learning model requires fine tuning of its hyperparameters. Manual tuning approach is the traditional way to tune hyperparameters. In this approach, different combinations of hyperparameters are set and tested manually, but it is time consuming and inefficient in cases where there are many hyperparameters to evaluate (Yang and Shami, 2020). Alternatively, automatic tuning approaches, like grid search, random search, and Bayesian optimization, can be used. In Bayesian optimization approach, a hyperparameter search space is created manually first. Then, the algorithm selects a set of hyperparameters from the search space, trains the model, and returns a cross-validation score. This score is then used by the surrogate function to recommend new hyperparameter set to the objective function that would most likely result in a higher accuracy. This iterative process is repeated until the desired number of evaluations is achieved. Bayesian optimization usually requires less evaluations to find the optimal hyperparameter set that the other automatic tuning approaches (Snoek et al., 2012; Yang and Shami, 2020).

In this study, both trip production and trip distribution models, the data were split into training and testing subsets with 4:1 ratio, which is the most common split ratio used for machine learning models (Lever et al., 2016). Moreover, Bayesian optimization approach was used to obtain the optimal hyperparameter set for each algorithm. The search spaces for RF, Bagging, ANN, and DNN algorithms were created using the hyperparameters shown in Figure 7. While setting up the optimization function, we used Tree Parzen Estimator (TPE) algorithm for the surrogate function and the maximum iterations was set to 150. In each iteration, trained models were evaluated using 10-fold cross-validation. To find the hyperparameter set that returns the highest cross-validation score, the objective function was to minimize the negative accuracy score. The best hyperparameter set for each algorithm is presented in Appendix A. Moreover, the optimal architectures of the algorithms were applied on the testing set to find their prediction accuracy. These models were then compared in terms of their prediction accuracy.

SHAP (SHapley Additive exPlanations), a post-hoc model interpretability method, was employed to explain the results and capture the importance of the explanatory variables. SHAP, which was introduced by Lundberg and Lee (2017), is a game theoretic approach to explain the output of any machine learning model based on Shapley values. SHAP determines the average marginal contribution of each variable to the model outcomes for each observation (Kalatian and Farooq, 2021).

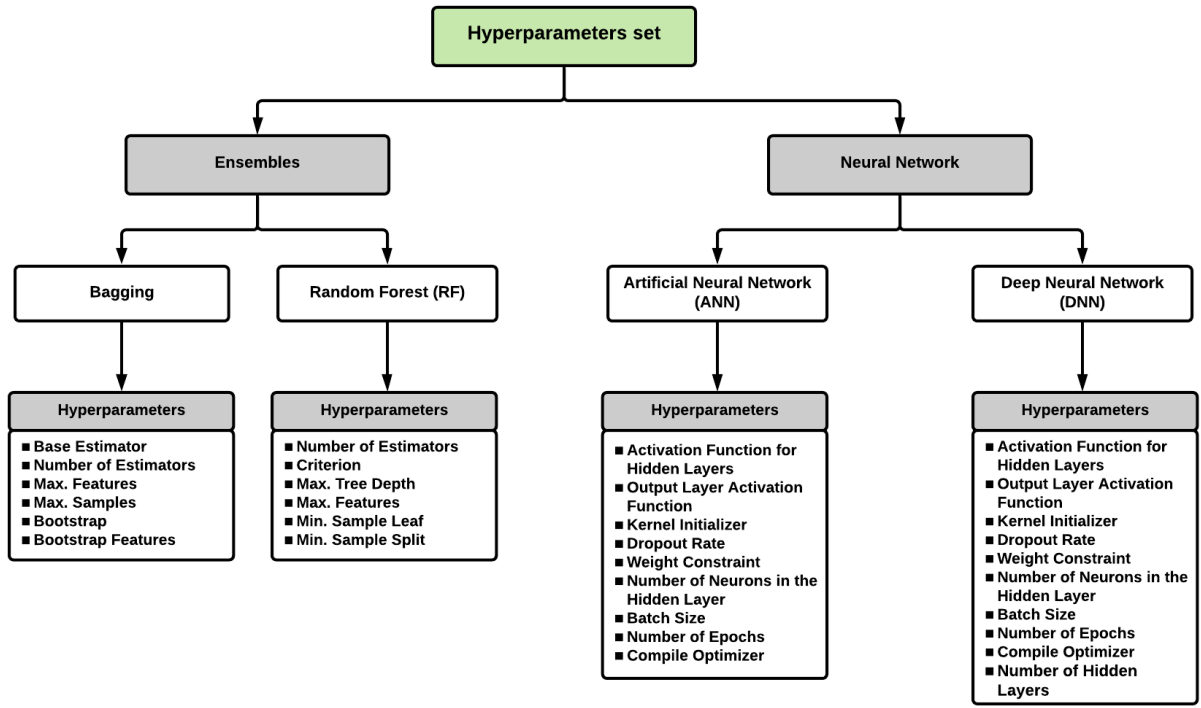

**Figure 7.** RF, Bagging, ANN, and DNN hyperparameters

## 5. Results

This section presents the results of the ODT daily trip production and distribution models.

5.1. ODT Daily Trip Production Model

The ODT daily trip production model was developed at DA level using RF, Bagging, ANN, and DNN algorithms. Here, we compare the performance of the four algorithms first, then we discuss the impotence and the impact of the explanatory variables on the model's outcome.

5.1.1. Model Comparison

Figure 8 compares the performance of RF, Bagging, ANN, and DNN algorithms in predicting the ODT daily trip production levels. It is observed that, in all algorithms, the low trip production level is the most accurately predicted level, followed by the high and the medium levels, respectively. The highest predictive accuracy for the low trip production level is obtained using DNN algorithm at 74%. While for the medium trip production level, the highest predictive accuracy is found using Bagging algorithm at 53%. On the other hand, the highest predictive accuracy for the high trip production level is 70%, achieved using ANN algorithm. Altogether, Bagging algorithm can best describe the ODT daily trip production levels with an overall prediction accuracy of 64%. The confusion matrix on the testing set for Bagging algorithm are presented in Table 2.

Moreover, it can be noticed that the model prediction accuracy is considerably lower for the medium trip production level compared to both low and high levels. This is mainly due to the fact that almost 45% of the observations for the medium trip production level are right on the lower boundary limit with 2 trips per day (see Figure 5b). As a result, the model misclassified a significant number of medium level values as a low level (see Table 2). However, the prediction accuracy of the model is expected to be improved in future work by incorporating more explanatory variables, like weather and land-use data, using different clustering techniques, and including more observations.

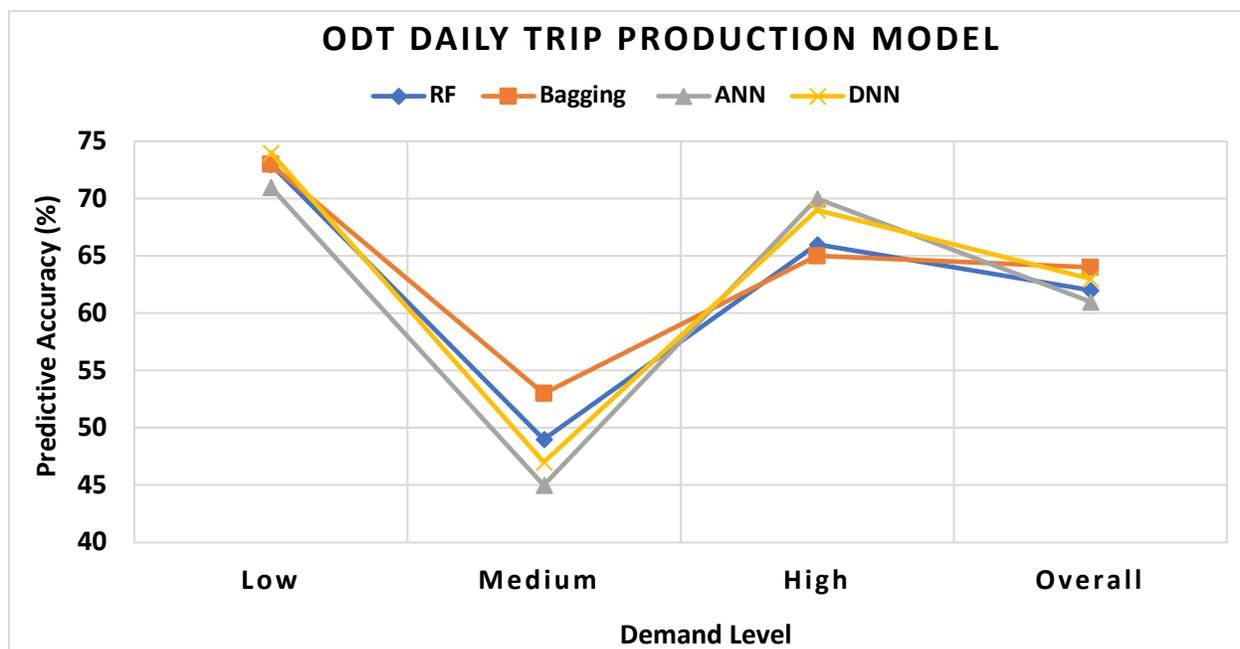

**Figure 8.** Comparing the performance of RF, Bagging, ANN, and DNN algorithms

**Table 2.** Trip production model testing confusion matrix

| | Confusion Matrix | Low Demand | Medium Demand | High Demand | Total | Accuracy (%) |
|---|---|---|---|---|---|---|
| | | | Predicted | | | |
| Actual | Low Demand | 166 | 54 | 6 | 226 | 73 |
| | Medium Demand | 78 | 114 | 24 | 216 | 53 |
| | High Demand | 1 | 29 | 55 | 85 | 65 |
| | Total | 245 | 197 | 85 | 527 | 64 |

5.1.2. Variable Importance and Effects

We used SHAP analysis technique to interpret the predictions of the Bagging model, best trip production model, and to analyze the importance of the explanatory variables. Figure 9 presents the SHAP results for the variable importance. It is seen that the population density is the most important variable in the model, followed by the percentage of males, median income, and working age percentage variables, respectively. These variables provide more than 70% of the model's interpretation. In contrast, the month-of-year and day-of-week variables are the least important variables in the model. These findings reveal the existence of a high spatial heterogeneity in the ODT trip production levels across the dissemination areas.

Figure 10, on the other hand, shows the effects of each variable on the model's outcome. Each point in Figure 10 represents a Shapley value for a variable and an observation. The variable determines its position on the y-axis, the Shapley value determines its position on the x-axis, and the value of the variable determines its colour. It can be noticed that the low population density values as well as and the high median income values increase the predicted trip production level. These findings are expected, as most users use the ODT service in Belleville to return home from work or shopping. Therefore, we can conclude that higher trip production levels are expected in the neighbourhoods with commercial/industrial land-use types than those with residential land-use type. It is also observed that the percentage of males, working age percentages, and the average household size variables have, in general, a high and positive effect on the trip production level. These findings are most likely the result of the home-to-work or home-to-shop trips. Hence, dissemination areas with higher percentage of males, working age percentages, and average household size are more likely to have higher trip production levels.

Moreover, the daily hours of operation variable has a positive correlation with the trip production level. This implies that the daily trip production level is more likely to be higher on weekends, which is due to the fact that the ODT service has higher hours of operation on weekends. On the other hand, the day-of-week variable is negatively correlated with the trip production levels. This indicates that the trip production level is more likely to be higher on weekends. Similarly, low month-of-year values increase the predicted trip production level, which indicates a higher trip production level during the first four months of the operation. However, as mentioned at the beginning of this section, both day-of-week and month-of-year variables are the least contributors to the predictions. To get more insights on the spread and variation of SHAP values with the variables, see the SHAP dependency plots presented in Appendix B.

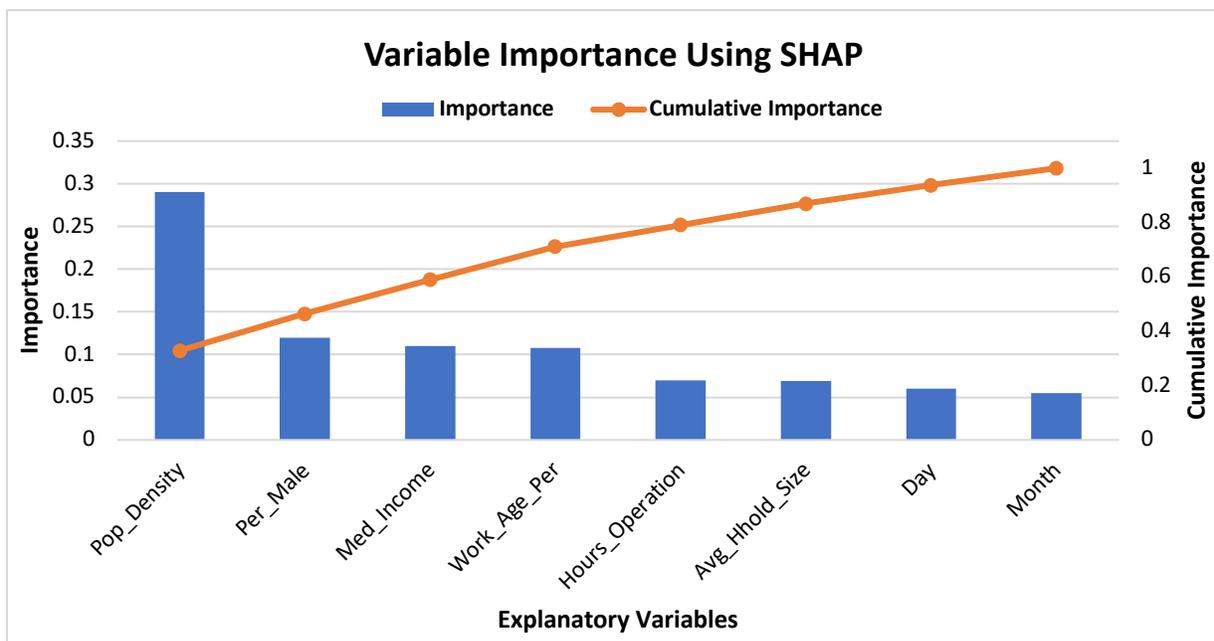

**Figure 9.** SHAP variable importance

5.2. ODT Daily Trip Distribution Model

The ODT daily trip distribution model was also developed at DA level using RF, Bagging, ANN, and DNN algorithms. This section compares the performance of the four algorithms and discusses the impotence and the impact of the explanatory variables on the model's outcome.

5.2.1. Model Comparison

The performance of RF, Bagging, ANN, and DNN algorithms in predicting the ODT daily trip distribution levels are presented in Figure 11. It is observed that all algorithms have a remarkably close performance over the daily trip distribution levels. However, the highest prediction accuracy for the low trip distribution level is obtained using DNN algorithm at 100%. On the other hand, the highest prediction accuracy for both medium and high trip distribution levels are found using RF algorithm at 65% and 63%, respectively. In general, RF algorithm can best explain the ODT daily trip distribution levels with an overall accuracy of 72%. Table 3 shows the confusion matrix on the testing set for RF algorithm.

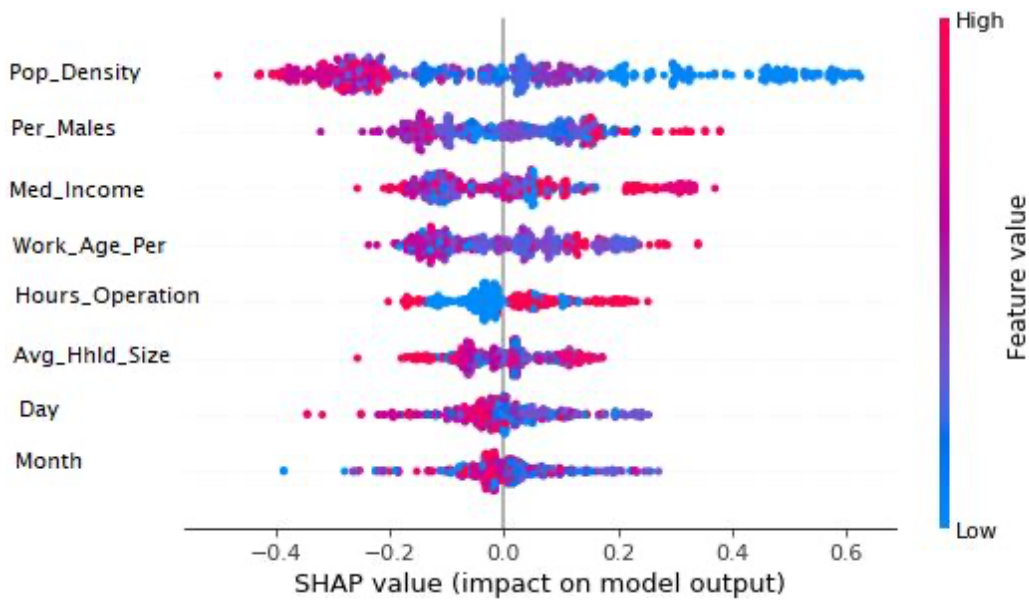

**Figure 10.** SHAP summary plot

Furthermore, it can be noticed that the low trip distribution model has a considerably higher prediction accuracy than medium and high levels. This is mainly due to the close proximity of the boundary set values between the medium and high trip distribution levels. As mentioned in the previous section, we intend to improve the prediction accuracy of both trip production and distribution models by incorporating more explanatory variables, using different clustering techniques, and including more including more observations.

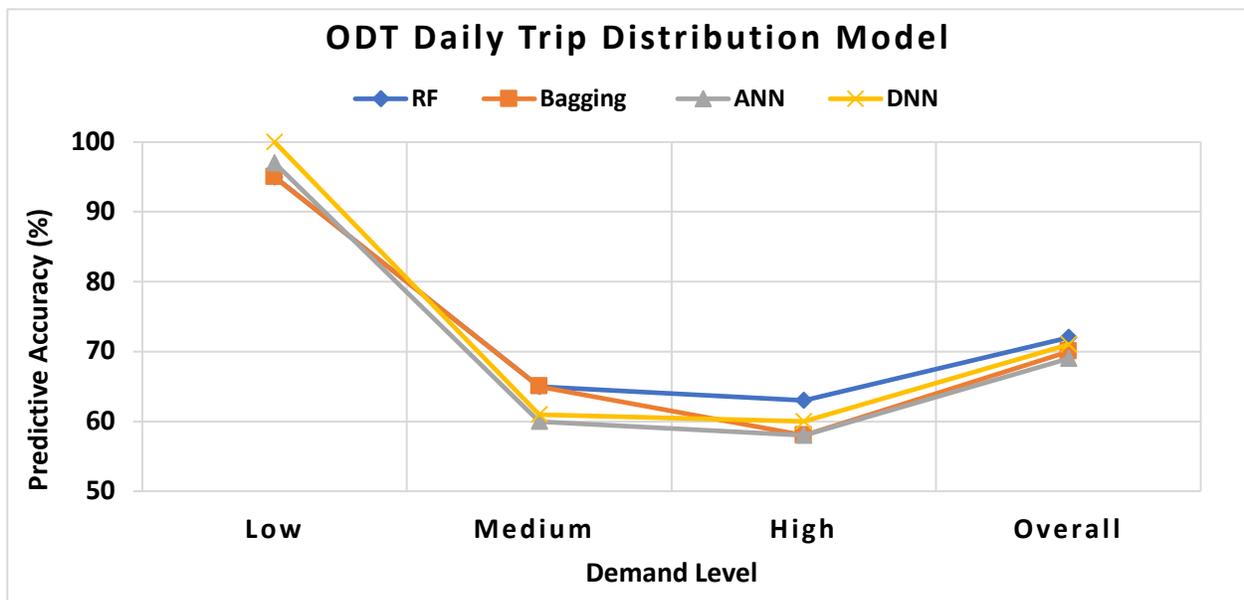

**Figure 11.** Comparing the performance of RF, Bagging, ANN, and DNN

**Table 3.** Trip distribution model testing confusion matrix

| Confusion Matrix | Predicted | | | Total | Accuracy (%) |
|---|---|---|---|---|---|
| | Low Demand | Medium Demand | High Demand | | |

|  | | | | | | |
|---|---|---|---|---|---|---|
| Actual | Low Demand | 107 | 3 | 3 | 113 | 95 |
| | Medium Demand | 32 | 109 | 27 | 168 | 65 |
| | High Demand | 14 | 44 | 98 | 156 | 63 |
| | Total | 153 | 156 | 128 | 437 | 72 |

5.2.2. Variable Importance and Effects

SHAP analysis technique was used to understand the outcome of the RF model, best trip distribution model, and to analyze the importance of the explanatory variables. The relative importance of the explanatory variables is presented in Figure 12. It is observed that the demographic characteristics of the trip destination are the most important variables in model. These characteristics can explain more than 86% of the model's predictions. However, the population density has the highest importance among the destination's demographic characteristics, followed in order by the median income and the average household size. These findings are in line with the Gravity model concept, which states that the number of trips between two zones is proportional to the attractiveness of the destination area (Pourebrahim et al., 2019). Moreover, the month-of-year variable has the highest importance among the other trip characteristics. On the other hand, the daily hours of operation, percentage of males, and the day-of-week variables are the least important variables in the model.

Figure 13 shows the effects of each variable on the daily trip distribution levels. Besides, the SHAP dependency plots for all variables are presented in Appendix C. It is observed that the population density, percentage of males, and working age percentage characteristics of the trip destination have a positive effect on the trip distribution levels. The lower percentage of males, higher median income, and lower working age percentages at the trip origin increase the predicted trip distribution level. These findings are not surprising, since the ODT service in Belleville is a night service and mainly used to return home from work or shopping. Hence, higher trip distribution levels are expected between dissemination areas with commercial/industrial land-use type and dissemination areas with high-density residential land-use.

Moreover, it can be noticed that higher median income as well as lower average-household size at the trip destination increase the predicted trip distribution level. While higher population density and average household size at the trip origin increase the predicted trip distribution level. This pattern is most likely the result of home-to-work or home-to-shop trips made by a second type of users. Therefore, this finding confirms that the trip distribution level is most likely to be higher between dissemination areas with multi-family residential land-use and dissemination areas with commercial/industrial land-use.

It is observed that the lower month-of-year values increase the predicted trip distribution level. This indicates a higher trip distribution level in the months December and January, which is probably due to the holiday seasons. Similarly, the day-of-week variable is negatively correlated with the trip distribution levels, indicating a higher trip distribution level on weekends. The daily hours of operation variable has a positive correlation with the trip production level. This confirms that the trip distribution level is most likely to be higher on weekends.

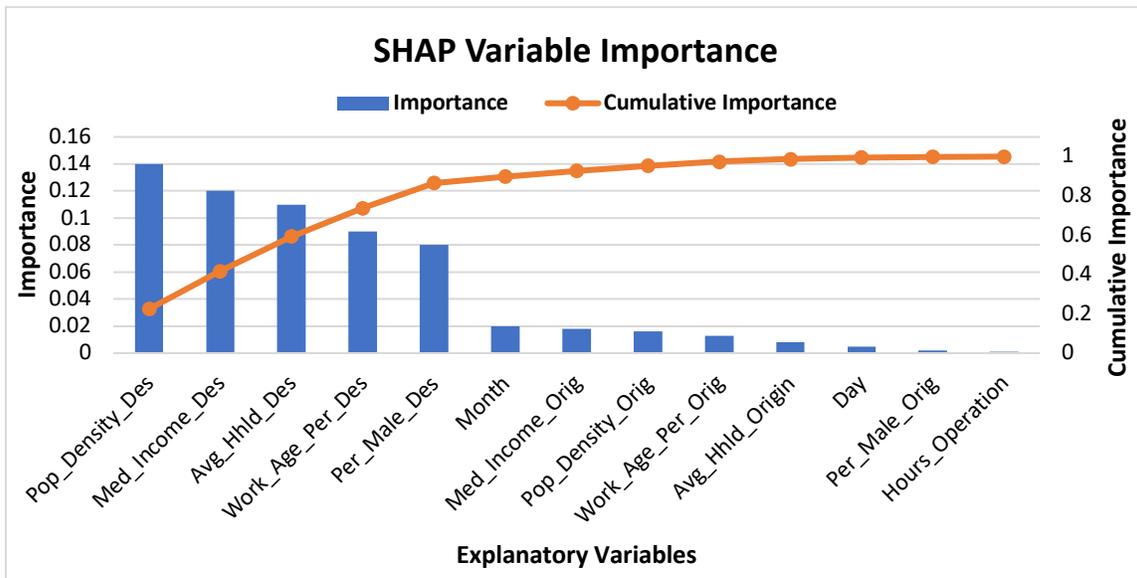

**Figure 12.** SHAP variables importance

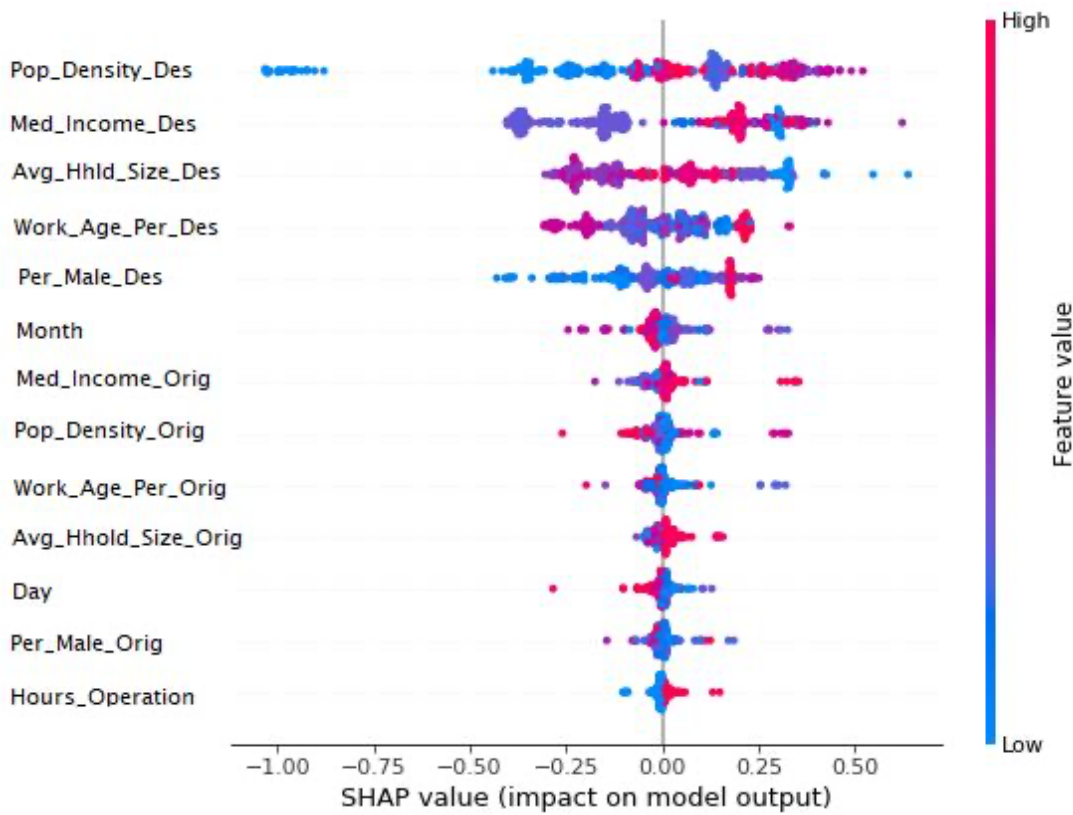

**Figure 13.** SHAP summary plot

## 6. Discussion and Policy Recommendations

This study modelled the daily trip production and distribution levels for Belleville's ODT at DA level. This work also discussed the importance and the effects of the explanatory variables on the ODT daily trip production and distribution levels. In this section, we compare our findings with the previous literature and provide useful recommendations to transit operators for sustainable planning, design, and operation of new and ongoing ODT projects.

## 6.1. Discussion

Aggregate and disaggregate travel demand models are complementary to help decision makers in making informed transportation planning decisions. Aggregate demand models can explain the spatial variations in demand between the zones, while disaggregate models can explain the variations in the usage frequency among individuals/households. The spatial resolution used in the previous demand modelling studies depended mainly on the purpose of the models and the level at which the modelling data was collected. Moreover, the demand models were developed at different temporal resolutions depending on the demand level of the service. In this study, to capture the spatial variations in the ODT demand among Belleville's zones and account for the low demand for the service, the trip production and distribution models were developed at a spatial scale of DA and a temporal scale of one day.

For the development of the trip production and distribution models, two main approaches have been adopted in the literature: machine learning algorithms and traditional statistical methods. In the current study, both models were developed using machine learning algorithms as they are more flexible, more accurate, and require fewer parameters to be estimated when compared with the traditional statistical methods (Pourebrahim et al., 2019; Tillema et al., 2006). Furthermore, unlike the previous demand modelling studies, which used regression algorithms to model the demand of ridesourcing services [for instance, Ye, (2019), Yan et al. (2020), and Li et al. (2017)], the demand of Belleville's ODT service was modelled using classification algorithms. This is mainly due to the very low daily demand values of the ODT service compared with the demand of ridesourcing services.

The results of this study reinforced previous findings in the existing literature that indicates the demand of on-demand shared mobility services is significantly affected by the land-use and the demographic characteristics of the service area (Jain et al., 2017; Ghaffar et al., 2020; Marquet, 2020; Yan et al., 2020). Our findings are also consistent with the study conducted by Sanaullah et al. (2021) which revealed that dissemination areas with commercial/industrial land-use type are likely to have higher ODT trip production levels than those with residential land-use. However, the relative importance of the demographic characteristics and the impact of the trip characteristics (daily hours of operation and trip date) on the ODT services have not been investigated in the literature before. Moreover, in line with previous studies, we found ensemble methods to have better performance than neural networks (Hsieh et al., 2011; Han et al., 2018; Rodriguez-Galiano et al., 2015; Pourebrahim et al., 2019).

Furthermore, the results revealed some key differences between on-demand dedicated fleet and crowdsourced mobility services in terms of user's characteristics and trip purposes. According to Marquet (2020) and Ghaffar et al. (2020), areas with higher population density and income tend to have higher ridesourcing trips. This is because ridesourcing services are mainly used to engage in non-work-related activities (Young and Farber, 2019). While in the current study, we found areas with lower population density and higher median income are associated with higher ODT trips, because most of the ODT users in Belleville use the service to return home from work or shopping. Therefore, we can conclude that ridesourcing users are mainly high-income passengers and the service is mostly used for no-work-related activities. On the other hand, most of Belleville's ODT users are low-income travellers who use the service to return home from work.

The City of Belleville has a low population density and an average median income that is a bit lower than the national average. It is the largest urban centre in Quinte region and comprises mainly of commercial/industrial neighbourhoods in the city centre surrounded by residential neighbourhoods. While this study is specific to the City of Belleville, the models and findings have the potential to be transferred to the other ODT projects in the areas that have the same community structure as in Belleville.

The main limitation of this work is the small dataset size used in modelling, which affected the prediction accuracy of the developed models. However, their accuracy is expected to be improved in future work by incorporating more explanatory variables, using different clustering techniques, and including more observations.

6.2. Policy Recommendations

There are several recommendations that can be derived from our results. These recommendations are of use to transit operators for sustainable planning, design, and operation of new as well as ongoing ODT projects. We found that dissemination areas with commercial/industrial land-use type are more likely to have higher daily trip production levels. Besides, higher daily trip distribution levels are foreseen between dissemination areas with multi-family residential land-use and dissemination areas with commercial/industrial land-use. In light of these findings, low-demand areas consisting of zones with commercial/industrial land-use as well as zones with multi-family residential land-use are the most favorable areas for the implementation of ODT services. Moreover, our results revealed the existence of a high spatial heterogeneity in the ODT trip production and distribution levels across the dissemination areas. To strike a balance between the performance of the service and its coverage, transit operators should consider modifying the service area by excluding low demand neighbourhoods. Furthermore, we provide the following suggestions to transit operators to enhance the performance of the service:

- Transit operators should continuously locate their empty vehicles in the neighbourhoods with commercial/industrial land-use as well as neighbourhoods with higher percentage of males, working age percentages, and average household size. This can help minimizing the vehicle miles travelled, total emissions, and users' waiting times.
- Transit operators should consider using the spatio-temporal demand models obtained in this work to continuously update the operating fleet size. Although these models may not be suitable for other ODT projects, they can provide key insights about the factors affecting the demand.
- The highest trip distribution level is mainly between neighbourhoods with multi-family residential land-use and neighbourhoods with commercial/industrial land-use. Therefore, we suggest converting the current ODT service into a MOD transit service, by using the current 40-ft buses to provide a fixed-route service between these neighbourhoods and offering ridesourcing services in the lower demand areas. However, a simulation-based study is needed to verify the feasibility and the validity of this suggestion.

**7. Conclusion**

This study modelled the daily trip production and distribution levels for Belleville's ODT at DA level. Previous literature revealed that machine learning algorithms are more flexible, more accurate, and require fewer parameters to be estimated compared with traditional statistical methods, like gravity model (Pourebrahim et al., 2019; Tillema et al., 2006). Hence, we used four machine learning algorithms for the development of the trip production and distribution models, namely Random Forest (RF), Bagging, Artificial Neural Network (ANN), and Deep Neural Network (DNN). The data used in the modelling process were acquired from Belleville's ODT operational data and 2016 census data. Bayesian optimization approach was used to find the optimal architecture of the adopted algorithms. Moreover, SHAP was employed to interpret the predictions and analyze the importance of the explanatory variables.

The results revealed that Bagging algorithm outperformed the other algorithms in predicting the ODT daily trip production levels with an overall prediction accuracy of 64%. On the other hand, RF algorithm outperformed the other algorithms in predicting the ODT daily trip distribution levels with an overall prediction accuracy of 72%. Moreover, the results revealed that the land-use type is the most important variable in the ODT trip production model, followed by the percentage of males, and working age percentage variables, respectively. While the month-of-year and the day-of-week variables are the least important variables in the trip production model. On the other hand, the demographic characteristics of the trip destination are the most important variables in the ODT trip distribution model. Among them, the population density has the highest importance, followed in order by the median income and the average household size. Furthermore, the results showed that dissemination areas with commercial/industrial land-use type are likely to have higher ODT trip production levels than those with residential land-use. The results also showed that higher trip distribution levels are expected between dissemination areas with commercial/industrial land-use type and dissemination areas with high-density residential land-use.

We provide useful recommendations for sustainable planning, design, and operation of new and ongoing ODT projects. Our findings suggest that the most favourable areas for the implementation of ODT services are low-demand areas consisting of zones with commercial/industrial land-use as well as zones with multi-family residential land-use. Moreover, the performance of ODT services can be further enhanced by (a) locating idle vehicles in the neighbourhoods with commercial/industrial land-use and (b) continuously updating the operating fleet size based on the spatio-temporal demand models obtained in this work. The main limitation of this work is the small dataset size used in modelling, which represents data collected from September 2018 to May 2019. The spatial and temporal transferability of the models was also not considered, which could be a future direction. We also intend to improve the prediction accuracy of both trip production and distribution models by incorporating more explanatory variables, using different clustering techniques, and including more including more observations. We also intend to investigate public transit user's preference between FRT and ODT services.

**ACKNOWLEDGEMENTS**

This research was funded by the Canadian Urban Transit Research & Innovation Consortium (CUTRIC) and by Ryerson University.

**Appendix A. Optimal Hyperparameter set**

Table A4 presents the optimal hyperparameter set for the adopted algorithms.

**Table A4.** Best hyperparameters set

| Algorithm | Hyperparameters | Best Values | |
| --- | --- | --- | --- |
| | | Trip Production Model | Trip Distribution Model |
| Random Forest (RF) | Number of Estimators | 150 | 50 |
| | Criterion | Gini | Entropy |
| | Max. Tree Depth | 150 | 40 |
| | Max. Features | Square root (Number of features) | Square root (Number of features) |
| | Min. Sample Leaf | Ceil (0.005362 x Sample size) | Ceil (0.0036382 x Sample size) |
| | Min. Sample Split | Ceil (0.013786 x Sample size) | Ceil (0.00452197 x Sample size) |
| Bagging | Base Estimator | KNeighborsClassifier | RandomForestClassifier |
| | Number of Estimators | 70 | 200 |
| | Max. Features | (0.513365 x Number of features) | (0.307943 x Number of features) |
| | Max. Samples | (0.9265818 x Sample size) | (0.210654 x Sample size) |
| | Bootstrap | True | False |
| | Bootstrap Features | False | False |
| Artificial Neural Network (ANN) | Activation Function for Hidden Layers | Hyperbolic Tangent (Tanh) | Rectified Linear Unit (ReLU) |
| | Output Layer Activation Function | Softmax | Softmax |
| | Kernel Initializer | Glorot Uniform | Uniform |
| | Dropout Rate | 0.0 | 0.0 |
| | Weight Constraint | 0.0 | 0.0 |
| | Number of Neurons in the Hidden Layer | 22 | 15 |
| | Batch Size | 10 | 10 |
| | Number of Epochs | 500 | 150 |
| | Compile Optimizer | Adadelta | Adam |
| Deep Neural Network (DNN) | Activation Function for Hidden Layers | Hyperbolic Tangent (Tanh) | Hyperbolic Tangent (Tanh) |
| | Output Layer Activation Function | Logistic (Sigmoid) | Logistic (Sigmoid) |
| | Kernel Initializer | Glorot Uniform | Normal |
| | Dropout Rate | 0.0 | 0.0 |
| | Weight Constraint | 0.0 | 0.0 |
| | Number of Neurons in the Hidden Layer | 20 | 15 |
| | Batch Size | 10 | 20 |
| | Number of Epochs | 100 | 300 |
| | Compile Optimizer | Adadelta | Adam |
| | Number of Hidden Layers | 6 | 4 |

**Appendix B. SHAP Dependency Plots for Trip Production Model**

Figure B13 and B14 present the SHAP dependency plots for the demographic characteristics and trip characteristics, respectively.

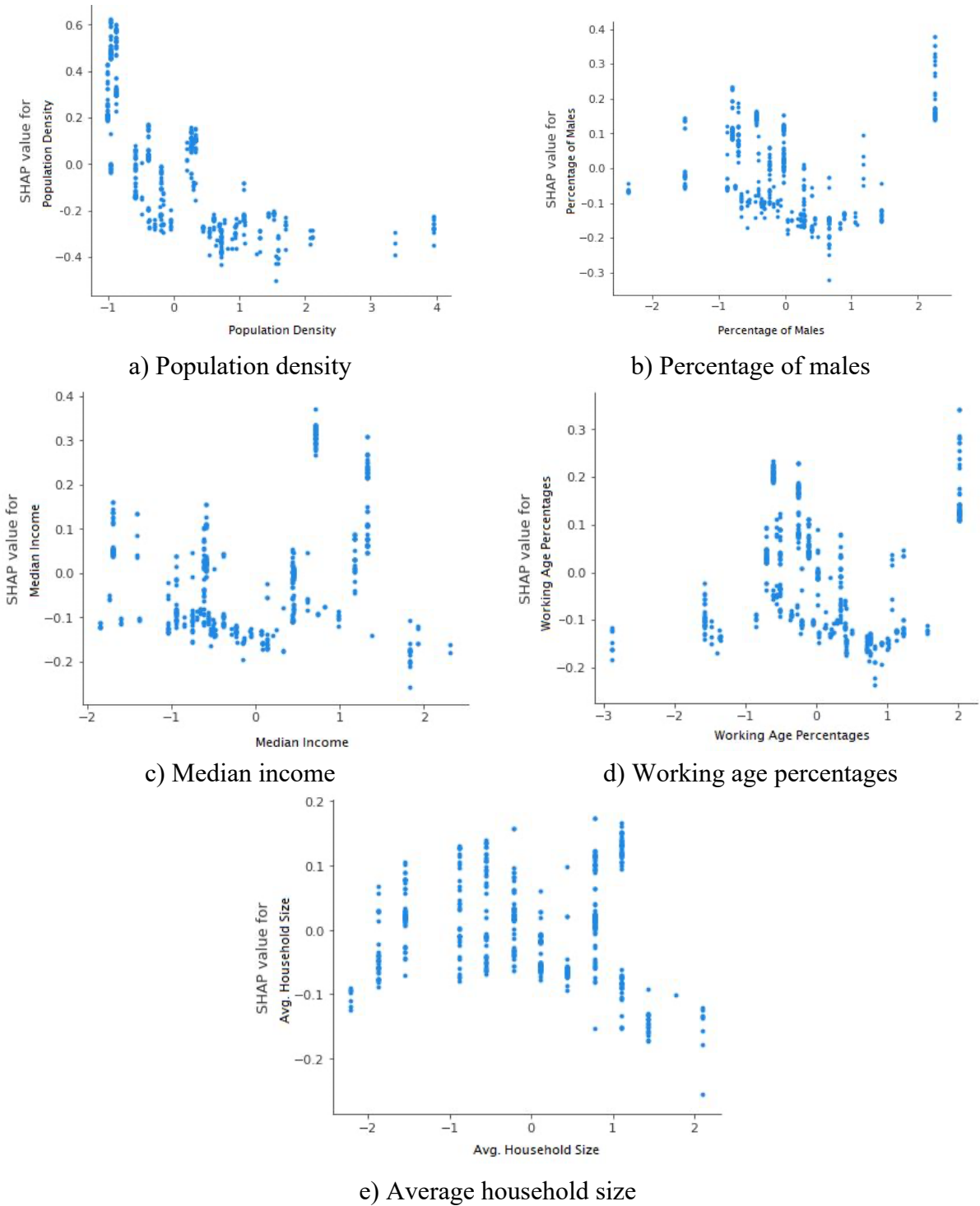

a) Population density

b) Percentage of males

c) Median income

d) Working age percentages

e) Average household size

**Figure B13.** SHAP dependency plots for the demographic characteristics

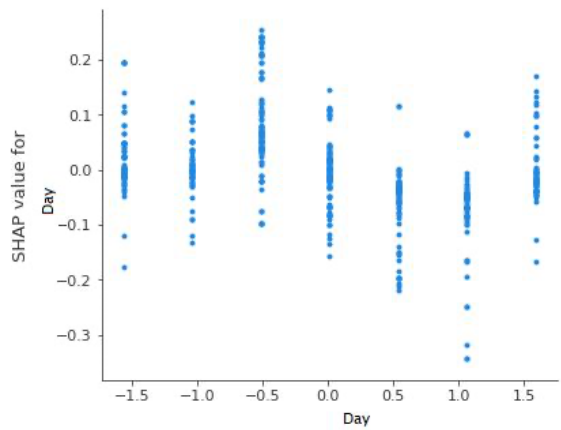

a) Day of the week

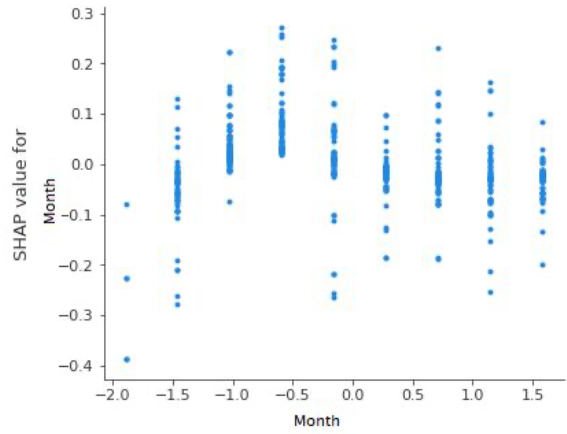

b) Month of the year

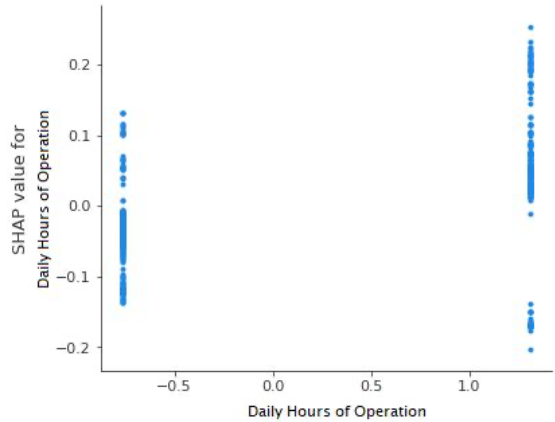

c) Daily hours of operation

**Figure B14.** SHAP dependency plots for the ODT trip characteristics

**Appendix C. SHAP Analysis Dependency Plots for Trip Distribution Model**

Figures C15 through C17 present SHAP dependency plots for the demographic characteristics of the trip destination, demographic characteristics of the trip origin, and the trip characteristics, respectively.

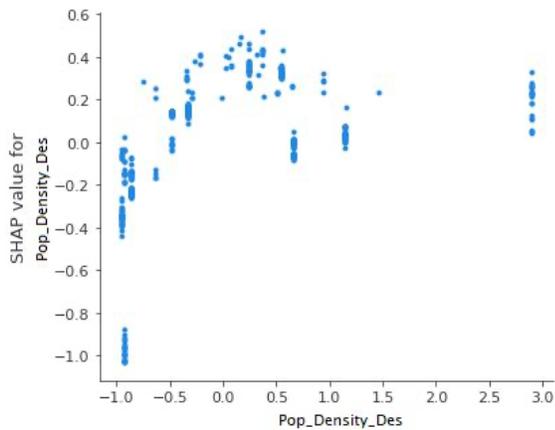

a) Population density

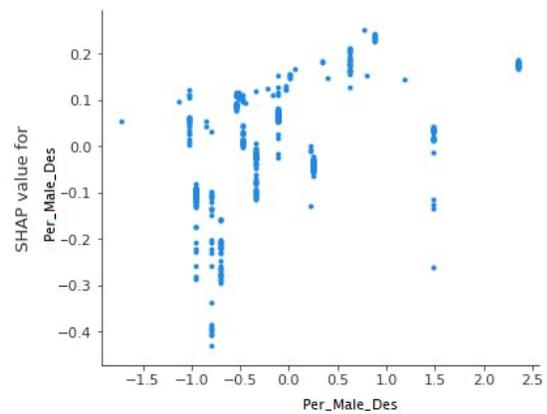

b) Percentage of males

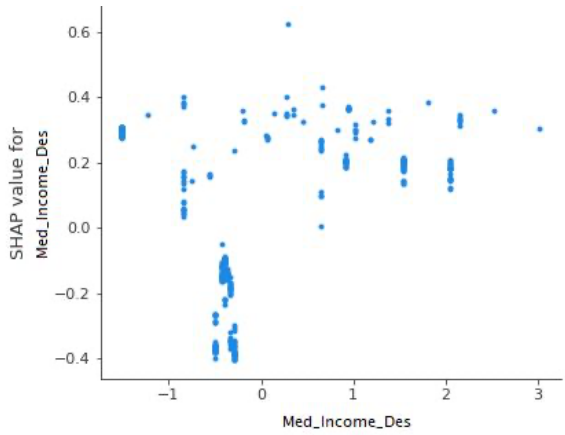
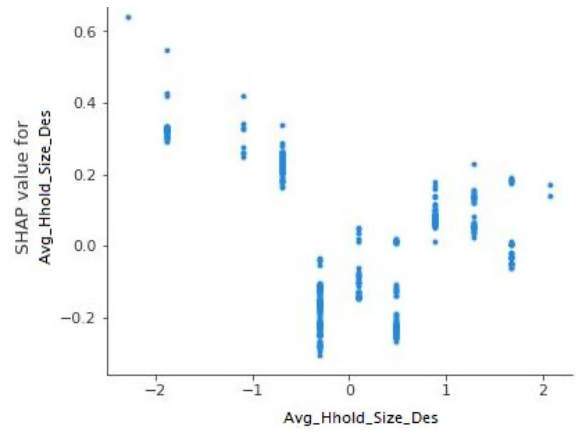

c) Median income
d) Average household size

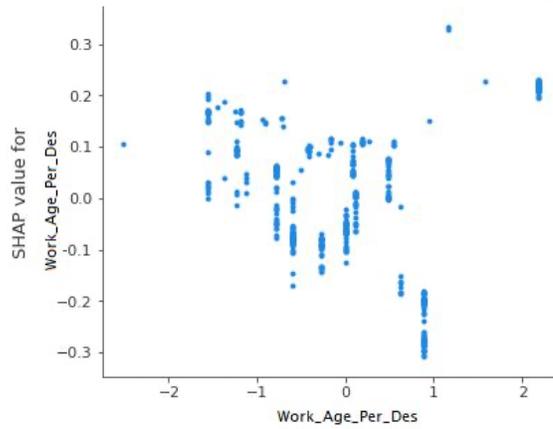

e) Working age percentages

**Figure C15.** SHAP dependency plots for the demographic characteristics of the trip destination

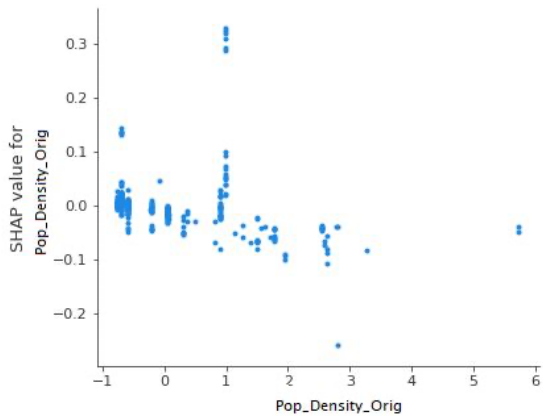
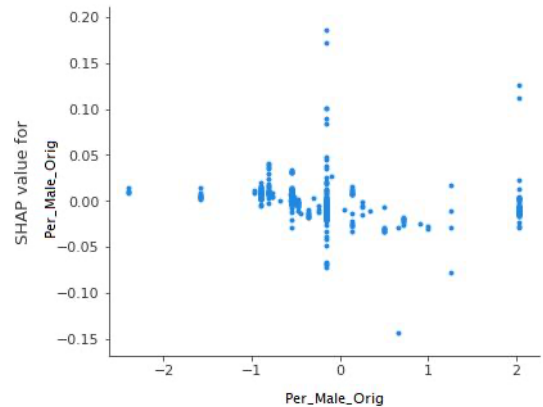

a) Population density
b) Percentage of males

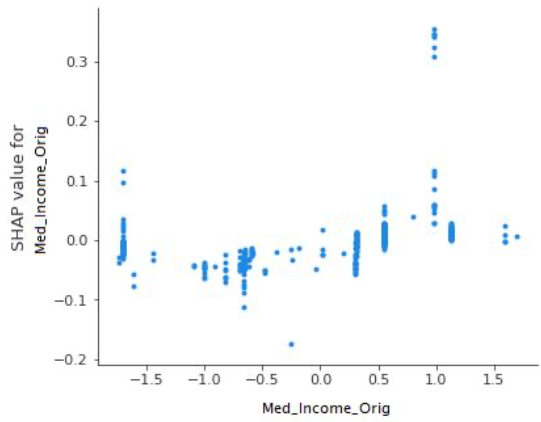
c) Median income

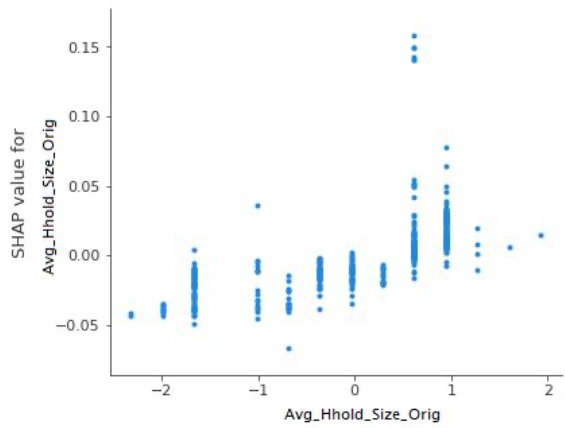
d) Average household size

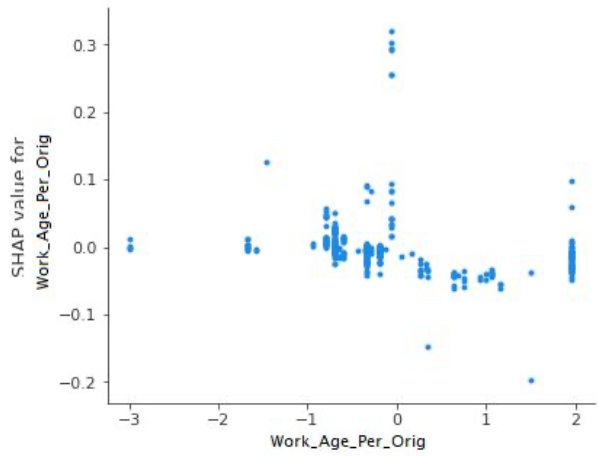
e) Working age percentages

**Figure C16.** SHAP dependency plots for the demographic characteristics of the trip origin

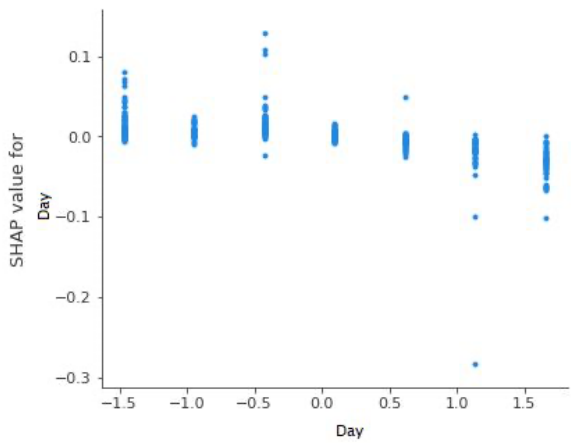
a) Day of the week

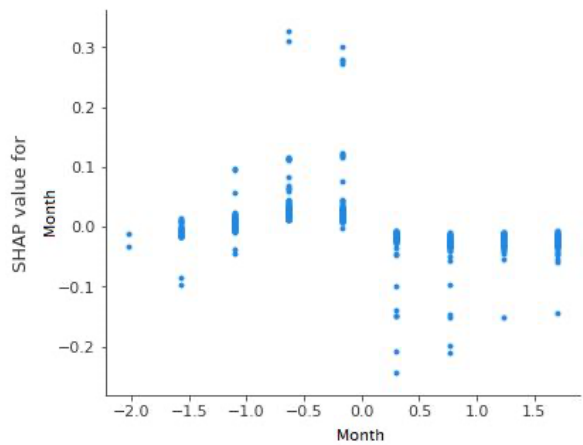
b) Month of the year

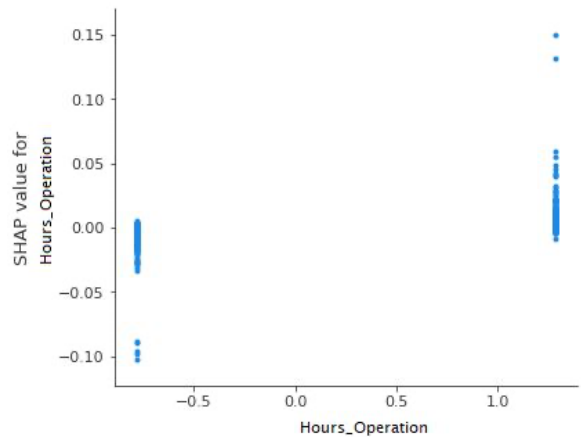

c) Daily hours of operation

**Figure C17.** SHAP dependency plots for the ODT trip characteristics